\documentclass[nofootinbib,showkeys,twocolumn,a4paper,10pt,aps,prd,preprintnumbers,superscriptaddress,floatfix]{revtex4-2}
\pdfoutput=1
\usepackage{amsmath,amssymb,graphicx,bm,psfrag,color,slashed,array,subfiles,cancel,braket,multirow,bigstrut,float}
\usepackage[version=4]{mhchem}
\usepackage[dvipsnames]{xcolor}
\usepackage{xspace} 
\usepackage{tikz}
\usepackage{lineno,hyperref}
\hypersetup{
    colorlinks=true,
    linkcolor=blue,
    filecolor=magenta,      
    urlcolor=cyan,
    pdftitle={Projected sensitivity of the ANUBIS detector to tau-coupled heavy neutral leptons},
    pdfpagemode=FullScreen,
    }
\usepackage{verbatim} 
\usepackage{siunitx} 
\DeclareSIUnit{\year}{y}

\usepackage{enumitem}
\setlist{itemsep=0mm}

\newcommand{\anubis}{\mbox{ANUBIS}\xspace}
\newcommand{\setanubis}{\mbox{\textsc{SET-ANUBIS}}\xspace}

\newcommand{\atlas}{\mbox{ATLAS}\xspace}

\newcommand{\codex}{\mbox{CODEX-b}\xspace}

\newcommand{\pythia}[1][]{\mbox{\textsc{Pythia#1}}\xspace}

\newcommand{\madspin}{\mbox{\textsc{MadSpin}}\xspace}
\newcommand{\madversion}[1]{MG5\_aMC@NLO #1}

\newcommand{\mm}{\ensuremath{\textnormal{mm}}\xspace}

\newcommand{\metre}{\ensuremath{\textnormal{m}}\xspace}

\newcommand{\GeV}{\ensuremath{\textnormal{GeV}}\xspace}
\newcommand{\TeV}{\ensuremath{\textnormal{TeV}}\xspace}

\let\iab=\invab

\newcommand{\met}{\ensuremath{E_T^\text{miss}}\xspace} 
\newcommand{\pt}{\ensuremath{p_T}\xspace} 

\let\BF=\br

\def\pmbanner{}
\begin{document}
\title{\pmbanner{\LARGE \texorpdfstring{Projected sensitivity of the \anubis detector to tau-coupled Heavy Neutral Leptons\\
\includegraphics[width=0.1\linewidth]{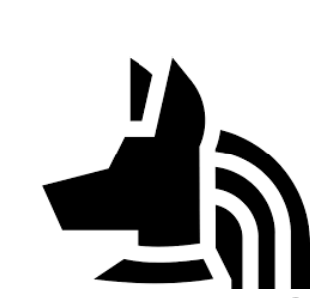}\\
{\large\textbf{ANUBIS Collaboration}}
\\[-0.2em]
{\normalsize\textit{E-mail: }\href{mailto:anubis-publications@cern.ch}{anubis-publications@cern.ch}}
\\[-0.2em]
}{Projected sensitivity of the ANUBIS detector to tau-coupled Heavy Neutral Leptons}}
}

\begin{abstract}
\textit{Abstract:}
Heavy Neutral Leptons (HNLs) provide a minimal and well-motivated extension of the Standard Model neutrino sector, and are naturally long-lived for small mixing angles between SM neutrinos and HNLs.
A previous ANUBIS study determined the projected sensitivity to a single Majorana HNL coupled exclusively to the electron or to the muon, corresponding to the BC6 and BC7 benchmark scenarios of the Physics Beyond Colliders initiative.
The present article extends that work to the complementary BC8 scenario, in which the HNL couples exclusively to the tau lepton.
As the tau lepton is itself unstable and because charged-current HNL decay modes become available only above the kinematic tau lepton production threshold, the production, decay and event-selection characteristics of this scenario cannot be obtained by rescaling the electron- or muon-coupled results in flavour.
The analysis is carried out end to end, from signal-sample generation through acceptance evaluation to sensitivity extraction, with the released SET-ANUBIS~v1.0.0 framework.
   \begin{center}
       \today 
   \end{center}
   \vspace{-0.5cm}
\end{abstract}

\author{Camilia Ben Messaouda}
\affiliation{Université Lyon 1, CNRS, IP2I, UMR 5822, Villeurbanne, France}
\affiliation{Formerly at: Cavendish Laboratory, University of Cambridge, Cambridge, United Kingdom}%
\author{Th\'eo Reymermier}
\affiliation{Université Lyon 1, CNRS, IP2I, UMR 5822, Villeurbanne, France}
\affiliation{Formerly at: Cavendish Laboratory, University of Cambridge, Cambridge, United Kingdom}
\author{Paul Swallow}
\affiliation{Cavendish Laboratory, University of Cambridge, Cambridge, United Kingdom}
\author{Oleg Brandt}
\affiliation{Cavendish Laboratory, University of Cambridge, Cambridge, United Kingdom}
\author{for the ANUBIS Collaboration}
\noaffiliation

\def\andname{\hspace*{-0.5em}}
\maketitle

\section{Introduction}
\label{sec:intro}
Long-Lived Particles (LLPs) arise in many extensions of the Standard Model (SM)~\cite{Antel:2023hkf} that address open questions such as the origin of neutrino masses~\cite{Bondarenko:2018ptm}, the nature of dark matter~\cite{Hall:2010jx,Cheung:2010gk,Zurek:2013wia}, and the baryon asymmetry of the Universe~\cite{Barry:2013nva}.
Their macroscopic decay lengths and weak couplings to ordinary matter motivate detector concepts that complement the general-purpose LHC experiments by instrumenting large decay volumes far from the interaction point.

\begin{figure}[ht]
    \centering
    \begin{tikzpicture}
        \node[inner sep=0] (image) at (0,0) {\includegraphics[width=0.75\linewidth]{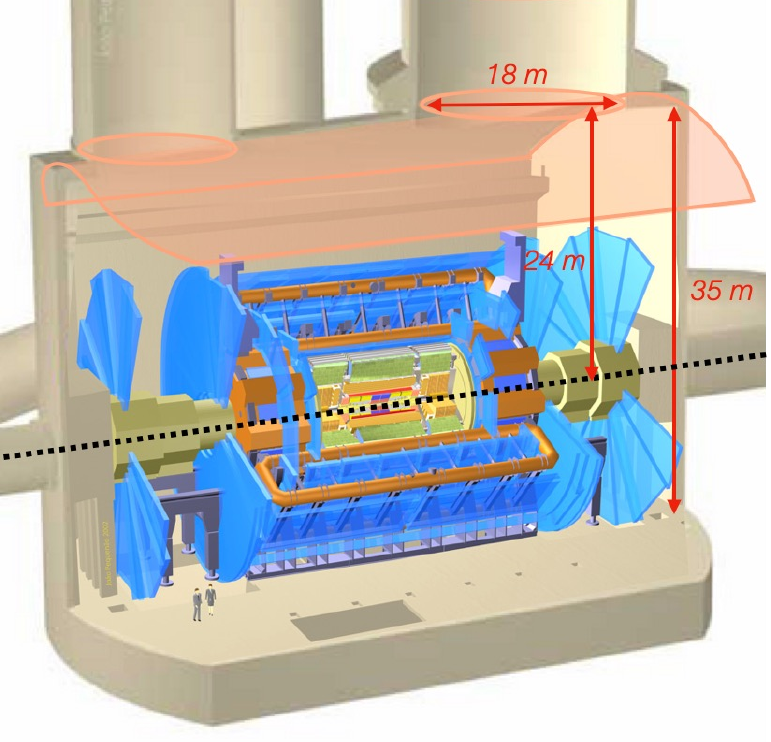}};
        \node[inner sep=0] at (1.1,2.9) {{\color{red}{\textbf{PX14}}}};
        \node[inner sep=0] at (-2,2.9) {{\color{red}{\textbf{PX16}}}};
    \end{tikzpicture}
    \caption{Layout of the \atlas underground cavern UX15 at IP1 of the LHC, including the access shafts PX14 and PX16. The area highlighted in orange illustrates the two RPC tracking layers of the \anubis detector.}
    \label{fig:SketchATLAS_undergroundCavern}
\end{figure}

The \anubis detector is proposed to instrument the ceiling of the \atlas UX15 cavern with Resistive Plate Chamber (RPC) tracking stations, as shown in Figure~\ref{fig:SketchATLAS_undergroundCavern}~\cite{Bauer:2019vqk,ANUBIS:2025sgg}.
Its transverse location and large air-filled decay volume provide sensitivity to electrically neutral LLPs that are produced at electroweak partonic centre-of-mass energies and decay at distances of order ten metres or more from the \atlas interaction point.
The detector concept, geometry, expected backgrounds, and baseline signal selections are motivated and described in detail in Ref.~\cite{ANUBIS:2025sgg}. 

Heavy Neutral Leptons (HNLs) are a particularly well-matched target for such a geometry, since only transverse detectors can access LLPs produced through comparatively heavy mediators.
A previous study by the \anubis Collaboration evaluated a minimal Majorana HNL coupled exclusively to the electron or to the muon, corresponding to the BC6 and BC7 benchmark cases of the Physics Beyond Colliders (PBC) initiative~\cite{ANUBIS:HNLemu2026,Beacham:2019nyx,Alemany:2019vsk}.
The present article completes this set of single-flavour benchmarks by considering BC8, in which the HNL mixes only with the tau neutrino.

The tau-coupled scenario is not a simple flavour relabelling of BC6 and BC7: it differs in production, decay, and event-selection features.
In charged-current production, the accompanying prompt charged lepton is an unstable tau lepton, so that its visible decay products and the associated neutrinos affect the event topology.
In HNL decays, final states containing an on-shell tau lepton only become kinematically accessible above the corresponding threshold, whereas neutral-current decays remain available below it.
Together, these effects alter the HNL total width, the visible branching fractions, the charged-track multiplicity, the missing transverse momentum, and the isolation efficiencies.

A second difference concerns the computational workflow.
The BC6 and BC7 analyses were carried out while \setanubis~\cite{erner2026setanubismodularpipelineanubis} was still under development.
The BC8 analysis is instead run end to end with the released \setanubis~v1.0.0 software~\cite{Reymermier:SETANUBISv100}, which provides a single reproducible pipeline for event generation, detector acceptance, event selection, category combination, and sensitivity extraction.
The two versions are comparable in functionality; the released one differs mainly in implementation detail and ease of use.
Only the ingredients specific to the tau benchmark and to this implementation are described below; the common detector and statistical assumptions are taken from Refs.~\cite{ANUBIS:2025sgg,ANUBIS:HNLemu2026}.

This article is organised as follows.
Section~\ref{sec:benchmark} summarises the BC8 HNL benchmark and the physical features specific to tau mixing.
Section~\ref{sec:analysis} describes the end-to-end \setanubis workflow used for the analysis.
The projected sensitivity is presented in Section~\ref{sec:results}, followed by the conclusions in Section~\ref{sec:conclusions}.

\section{Tau-coupled Heavy Neutral Lepton benchmark}
\label{sec:benchmark}
The model considered here is the same minimal Majorana HNL scenario used in Ref.~\cite{ANUBIS:HNLemu2026}.
A single right-handed neutral fermion, $N_1$, is added to the SM and interacts only through its mixing with the SM neutrinos.
After electroweak symmetry breaking, the relevant charged-current and neutral-current interactions read
\begin{align}
\label{eq:cc}
\mathcal{L}_{\text{HNL}}^W &= -\frac{g V_{1\alpha}}{\sqrt{2}} \left[ W^-_\mu(\overline{\ell}_{\alpha}^L \gamma^\mu N_1)\right]+\mathrm{h.c.}, \\ 
\label{eq:nc}
\mathcal{L}_{\text{HNL}}^Z&=-\frac{\sqrt{g^2 + g'^2}}{2}V_{1\alpha} Z_\mu\left(\overline{\nu}_{\alpha}^L \gamma^\mu N_1\right)+\mathrm{h.c.},
\end{align}
where $\alpha \in \{e,\mu,\tau\}$, $\nu_\alpha^L$ is the neutrino component of the SM left-handed lepton doublet, $V_{1\alpha}$ is the mixing parameter between $\nu_\alpha^L$ and the HNL, and $g$ and $g'$ are the weak isospin and weak hypercharge couplings, respectively.
An analogous Higgs interaction, proportional to $m_{N_1}V_{1\alpha}/v$, is also present, where $v$ is the Higgs vacuum expectation value.
The complete model definition, including the relation between the Yukawa coupling, the mass eigenstates and the mixing between SM neutrinos and HNLs is given in Ref.~\cite{ANUBIS:HNLemu2026}.

For the BC8 benchmark, the flavour structure is fixed to
\begin{equation}
|V_{1e}|:|V_{1\mu}|:|V_{1\tau}|=0:0:1,
\end{equation}
leaving the HNL mass, $m_{N_1}$, and the mixing magnitude, $|V_{1\tau}|^2$, as the two free parameters of the phenomenological study.
The Majorana nature of $N_1$ implies that charge-conjugate production and decay modes are included throughout.

\begin{figure}[ht]
    \centering
    \includegraphics[width=0.46\linewidth]{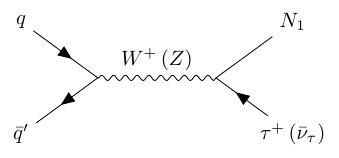}\hfill
    \includegraphics[width=0.48\linewidth]{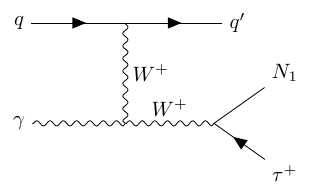}\\[1ex]
    \includegraphics[width=0.42\linewidth]{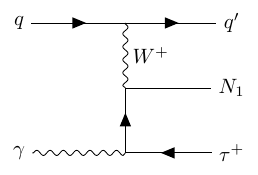}\hfill
    \includegraphics[width=0.48\linewidth]{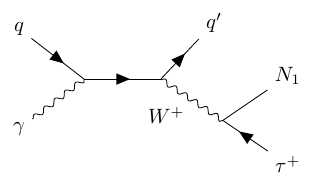}\\[1ex]
    \includegraphics[width=0.52\linewidth]{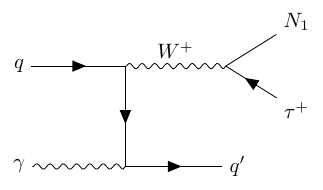}
    \caption{
    Direct electroweak HNL production modes considered in this study. The top-left diagram represents charged- and neutral-current Drell--Yan production; the remaining diagrams represent $W\gamma$ fusion.
    Charge-conjugate processes are implied.
    }
    \label{fig:production}
\end{figure}

The direct electroweak production modes retained from the previous study are
\begin{eqnarray*}
q\bar q'\to W^{*\pm}\to N_1\tau^{\pm} &:& \mathrm{CCDY},\\
q\bar q\to Z^*\to N_1\nu_{\tau}/\bar\nu_{\tau} &:& \mathrm{NCDY},\\
q\gamma\to q'\tau^{\pm}N_1 &:& W\gamma\ \mathrm{fusion}.
\end{eqnarray*}
Their representative Born-level diagrams are shown in Figure~\ref{fig:production}.
Hadronic production from heavy-flavour decays is not included due to their negligible acceptance on account of isolation requirements on LLP candidates imposed by the data-driven background estimate from Ref.~\cite{ANUBIS:2025sgg}. Hence ANUBIS’ reach for $m_{N_1}\lesssim 4~\GeV$ in this article is conservative.
This treatment is consistent with the BC6 and BC7 analysis, where this production mode was found to be negligible~\cite{ANUBIS:HNLemu2026}.

For $m_{N_1}\ll m_W$, the HNL decays through off-shell electroweak bosons.
Neutral-current decays of the form $N_1\to\nu_{\tau}f\bar f$ are kinematically allowed over the full mass range considered, and include visible leptonic and hadronic final states as well as the invisible three-neutrino channel.
Charged-current modes containing an on-shell tau lepton, such as ${N_1\to\tau^{\pm}q\bar q'}$ and ${N_1\to\tau^{\pm}\ell^{\mp}\nu_{\ell}}$, become accessible only above their kinematic thresholds.
The rapid change in available channels around $m_{N_1}\simeq m_{\tau}$ modifies both the proper lifetime and the visible branching fraction.
Above threshold, the subsequent tau lepton decay introduces additional neutrinos and can yield either a single track from a charged particle or a three-prong track system.

\section{Analysis with SET-ANUBIS}
\label{sec:analysis}
\setanubis is a modular framework for simulating LLP models in \anubis and for evaluating the corresponding acceptances and sensitivities~\cite{Reymermier:SETANUBISv100}.
Version~1.0.0 is used here for the complete BC8 workflow.
Model configuration, generator steering, HepMC event handling~\cite{Buckley:2019xhk}, particle propagation through the decay-volume, kinematic selections, acceptance aggregation, and sensitivity-contour extraction are all executed within the same released pipeline.
\setanubis was validated in the BC6 and BC7 study\cite{ANUBIS:HNLemu2026}, when parts of its architecture were still evolving; this is the first analysis to use the released v1.0.0.

\subsection{Signal generation}
\label{sec:generation}
Signal events are generated for $pp$ collisions at ${\sqrt{s}=14~\TeV}$ using \madversion{3.5.4}~\cite{Alwall:2014hca} with the NN23LO1 PDF set~\cite{Ball:2013hta} and the \textsc{SM\_HeavyN\_NLO} UFO implementation~\cite{Ruiz:2020cjx,Alva:2014gxa,Degrande:2016aje}.
The production channels are CCDY, NCDY, and $W\gamma$ fusion, as defined in Section~\ref{sec:benchmark}.
HNL decays are handled with \madspin~\cite{Artoisenet:2012st}, followed by parton showering, hadronisation, and unstable-particle decays with \pythia{8}~\cite{Bierlich:2022pfr}.
Tau lepton decays in particular are simulated explicitly, so that the charged particles and neutrinos entering the \anubis selection are those of the complete final state rather than of an undecayed tau lepton object.

Samples are organised by HNL production and decay mode.
The contribution of each category is reconstructed from the corresponding production cross-section, HNL branching fraction, and acceptance.
This treatment automatically assigns zero weight to kinematically forbidden on-shell-tau lepton decay modes and accounts for the change in the total HNL width across the tau lepton threshold.
It also avoids both the assumption of flavour universality for the acceptance and the application of a posteriori rescaling factors to the BC6 or BC7 samples.

The $(m_{N_1},|V_{1\tau}|^2)$ scan covers the same mass--mixing domain as the electron- and muon-coupled study, with additional sampling in the regions where the thresholds or the projected sensitivity vary rapidly.
The full scan definition, generator settings, random seeds, and sample bookkeeping are stored in the \setanubis analysis configuration accompanying this study.

\subsection{Acceptance and event selection}
\label{sec:selection}
The \anubis detector geometry and the baseline event selections follow those of Ref.~\cite{ANUBIS:HNLemu2026}, and are provided in Tables~\ref{tab:PartDefinition} and~\ref{tab:Selection} in Appendix~\ref{app:Selection}.
In summary: an accepted HNL must decay inside the air-filled fiducial volume between \atlas and the \anubis tracking stations.
At least two charged decay products must intersect at least two tracking layers each, so as to provide a reconstructable displaced-vertex candidate.
The baseline kinematic selection further requires the magnitude of the momentum vector to be ${|\vec p_{N_1}|>0.1~\GeV}$, so that the decay products are detectable with the RPCs; missing transverse momentum ${\met>30~\GeV}$; and an angular separation ${\Delta R>0.5}$ between the HNL direction and any energetic prompt jet or charged particle to reduce background contributions from $K_L$ and neutrons.
Jets are reconstructed with the anti-$k_t$ algorithm~\cite{Cacciari:2008gp} with radius parameter $R=0.4$, as in the preceding \anubis studies.

\begin{figure} [ht]
    \centering
    \includegraphics[width=1\linewidth]{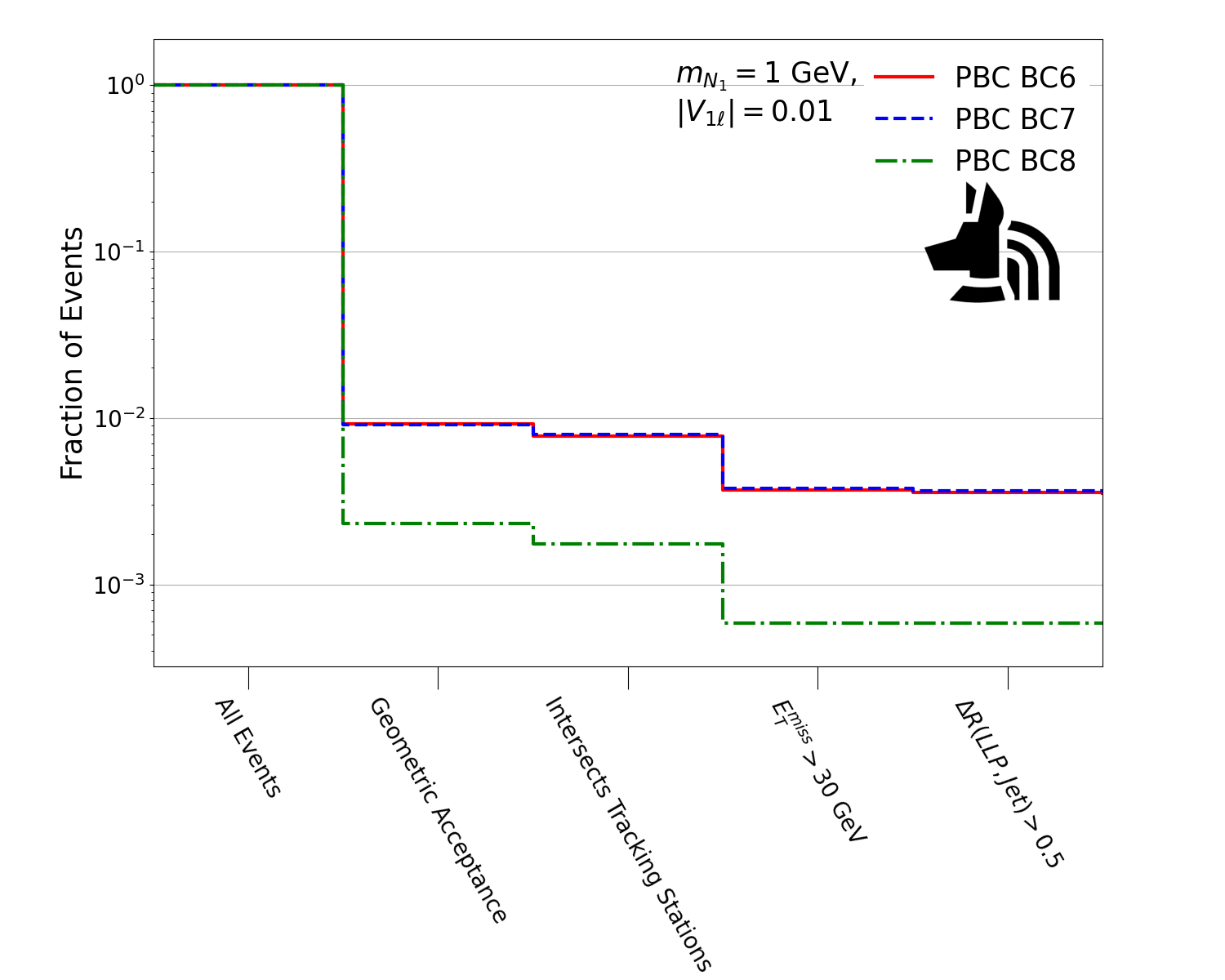}
    \caption{Fraction of remaining signal events after applying the successive selection requirements for the BC8 PBC scenario with  ${m_{N_1}=1~\GeV}$ and ${|V_{1\ell}|=0.01}$, in comparison to BC6 and BC7 from previous studies~\cite{ANUBIS:HNLemu2026}.}
    \label{fig:HNLcutflow}
\end{figure}

For BC8, all object definitions are applied after the prompt tau lepton decays have been simulated.
Tau-decay neutrinos contribute to the missing transverse momentum, while visible one-prong and three-prong decays contribute to the prompt charged-particle and jet collections used by the isolation requirements.
Likewise, when an HNL decay contains a tau lepton, the \anubis charged-track requirement is evaluated on the stable charged daughters of that tau.
This is the principal detector-level distinction from a direct $e/\mu\to\tau$ substitution in the previous analysis.
The effect of the successive selection requirements on BC8 is shown in Figure~\ref{fig:HNLcutflow}, with the corresponding BC6 and BC7 results shown for reference~\cite{ANUBIS:HNLemu2026}.
It also brings out a further distinction from those cases: a kinematic suppression at low HNL mass, amounting to roughly an order of magnitude in acceptance.

\subsection{Sensitivity calculation}
\label{sec:sensitivity}
The expected number of selected HNL decays is evaluated as a sum over production, $p$, and decay, $d$, modes,
\begin{align}
N_{\mathrm{HNL}}
&=\mathcal{L}\,\varepsilon_{\mathrm{sig}}
\sum_{p,d}\sigma_p(m_{N_1},|V_{1\tau}|^2)\,
\BF_d(m_{N_1})\nonumber\\
&\hspace{2.0cm}\times
\mathcal{A}_{p,d}(m_{N_1},|V_{1\tau}|^2),
\label{eq:yield}
\end{align}
where $\mathcal{A}_{p,d}$ is the geometrical and kinematic acceptance returned by \setanubis and $\BF_d$ is the branching fraction of the LLP's decay.
Following Ref.~\cite{ANUBIS:2025sgg}, an integrated luminosity of $\mathcal{L}=3~\iab$, as expected at the High-Luminosity LHC, is assumed, together with an \anubis signal detection efficiency of $\varepsilon_{\mathrm{sig}}=50\%$.
The latter corresponds to the worst case for detection at \anubis, namely displaced vertices with low track multiplicity~\cite{ANUBIS:2025sgg}.
The nominal 95\% confidence-level (CL) exclusion contour is defined by $N_{\mathrm{HNL}}=4$ selected events, which corresponds to an effectively background-free environment that can be achieved by exploiting the presence of a prompt lepton in the final state, and is consistent with the treatment in the BC6 and BC7 projections~\cite{ANUBIS:HNLemu2026}. It is assumed that these prompt leptons have trigger and selection efficiencies within \atlas of 100\%.

The CCDY and $W\gamma$ production modes contain a prompt tau lepton of which the visible decay products can be exploited by \atlas, whereas NCDY events contain an associated neutrino and are characterised primarily by missing transverse momentum. The background-free assumption is currently applied for all production modes.
A channel-dependent background analysis using the combined \atlas--\anubis event information lies beyond the scope of this projection; it can be incorporated in future \setanubis studies by replacing the event threshold in Equation~\eqref{eq:yield}.

\section{Results}
\label{sec:results}

\begin{figure}[ht]
    \centering
    \includegraphics[width=1\linewidth]{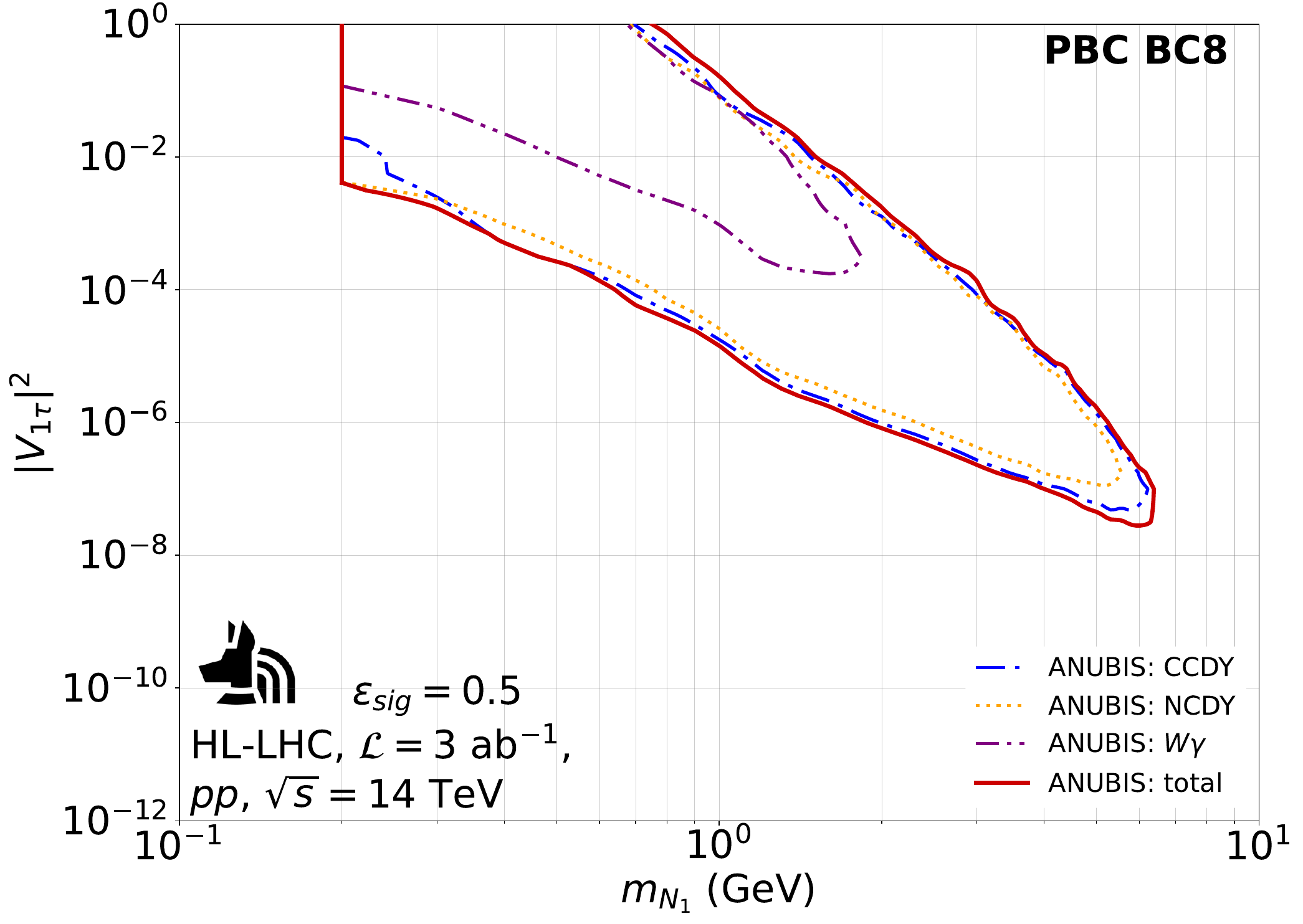}
    \caption{Projected exclusion sensitivity of the \anubis detector to a minimal Majorana HNL scenario that exclusively has $|V_{1\tau}|$ couplings, corresponding to the BC8 PBC scenario. 
    The combined sensitivity (solid line) is overlaid with the individual contributions from the different production modes (dashed/dotted lines).}
    \label{fig:ContourForProductionModes}
\end{figure}

The expected sensitivity to the tau-coupled Majorana HNL model, corresponding to the BC8 benchmark scenario of the PBC initiative, is presented in Figure~\ref{fig:ContourForProductionModes}.
The figure shows the exclusion contours obtained in the ${(m_{N_1},~|V_{1\tau}|^2)}$ parameter space for the dominant production modes, CCDY, NCDY and $W\gamma$ fusion, each combining all decay modes defined in Section~\ref{sec:benchmark}, together with the combined limit.

As in the electron- and muon-coupled cases of BC6 and BC7~\cite{ANUBIS:HNLemu2026}, the contours take a fin-like shape: the largest $m_{N_1}$ values are probed for the smallest couplings, while small $m_{N_1}$ is reached only at larger couplings.
This is because $c\tau\propto|V_{1\tau}|^{-2}$, so in order to remain within the geometric acceptance of \anubis a larger $m_{N_1}$ value is needed for smaller couplings. Also, at higher masses the decays are relatively prompt at a given coupling due to the kinematically accessible decay modes.
The CCDY production mode again dominates, reflecting the larger $W$ production cross-section relative to the other modes.

The tau-coupled HNL scenario, BC8, has a reduced reach relative to BC6 and BC7 because decay channels containing an on-shell tau lepton are kinematically accessible only for $m_{N_1}>m_{\tau}$.
A second difference from the electron- and muon-coupled cases is that every charged-current HNL decay mode now produces a tau lepton, which is itself unstable and decays predominantly hadronically.
This increases the charged-track multiplicity in the HNL decay, so a higher vertex-reconstruction efficiency is expected; the conservative $\varepsilon_{\rm sig}=50\%$ is nevertheless retained for consistency with BC6 and BC7.
Overall, the exclusion sensitivity for the BC8 scenario is found to be similar to that of BC6 and BC7, albeit slightly weaker in its $|V_{1\ell}|^2$ and $m_{N_1}$ reach. 
A direct comparison of exclusion sensitivities for BC6, BC7, and BC8 is provided in Figure~\ref{fig:AllHNLLimits} in Appendix~\ref{app:SummaryLimits}.

\begin{figure}[ht]
    \centering
    \includegraphics[width=1\linewidth]{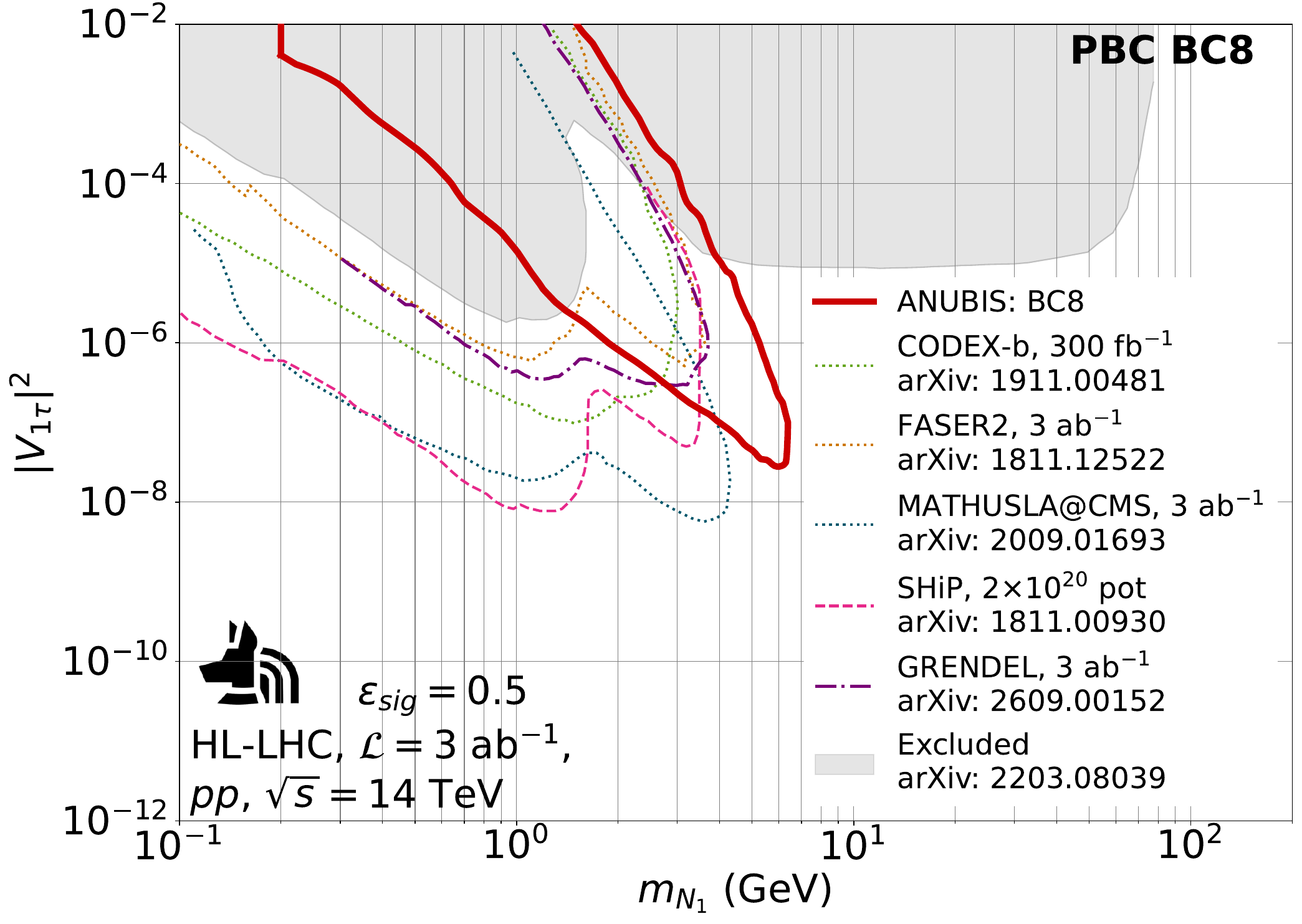}
    \caption{Projected exclusion sensitivity of the \anubis detector to a minimal Majorana HNL scenario that exclusively has $|V_{1\tau}|$ couplings, corresponding to the BC8 PBC scenario.
    Also shown are projected sensitivities and excluded regions from previous searches, summarised in Refs.~\cite{PBC:2025sny,Aielli:2019ivi,ATL-PHYS-PUB-2025-048}.}
    \label{fig:ContourWithExperiments}
\end{figure}

The projected \anubis limits are placed alongside existing constraints and the projections of other proposed experiments in Figure~\ref{fig:ContourWithExperiments}.
Below $m_{N_1}\approx4~\GeV$, \anubis covers parameter space that is complementary to the present constraints, part of which may also be reached by \codex~\cite{Aielli:2019ivi} and SHiP~\cite{SHiP:2021nfo}.
At larger masses, $m_{N_1}\gtrsim4~\GeV$, among the proposals considered only \anubis has sensitivity to small couplings, with a best sensitivity of ${|V_{1\tau}|^2=2.5\times10^{-8}}$ at $m_{N_1}=6.2~\GeV$.

\section{Conclusions}
\label{sec:conclusions}
Exploiting the full physics potential of the HL-LHC calls for a transverse experiment able to reach the region of LLP parameter space characterised by decay lengths of $\gtrsim10~\metre$ and partonic centre-of-mass energies at or above the electroweak scale.
\anubis is a proposed detector of exactly this kind.
 
The sensitivity of \anubis to the minimal Majorana HNL model in the BC8 benchmark scenario of the PBC working group~\cite{Alemany:2019vsk} has been evaluated with version~1.0.0 of the \setanubis framework.
The projected reach extends beyond that of existing searches over ${2\lesssim m_{N_1}\lesssim 7~\GeV}$ and ${10^{-8}\lesssim|V_{1\tau}|^2\lesssim10^{-5}}$, and is unique among the proposed experiments considered here for $m_{N_1}\gtrsim4~\GeV$, peaking at ${|V_{1\tau}|^2=2.5\times10^{-8}}$ for $m_{N_1}=6.2~\GeV$.
Of the three single-flavour PBC benchmarks, the tau-coupled case remains the least constrained, with existing limits on $|V_{1\tau}|^2$ considerably weaker than those on $|V_{1e}|^2$ and $|V_{1\mu}|^2$.
This makes the region particularly well motivated for a dedicated search with a transverse detector such as \anubis.

\section{Acknowledgements}
\label{sec:ack}
We are grateful to Benjamin Fuks for his advice concerning the HNL model and its MadGraph implementation.
We thank everyone who contributed to the development and validation of the \setanubis framework, and we acknowledge the support of UKRI (CrossRefID 100014013) through the Future Leaders Fellowship scheme under grant number MR/V023098/1.

\clearpage
\bibliographystyle{JHEP}
\bibliography{bibs/theory,bibs/general,bibs/anubis-standard-refs}

\clearpage
\appendix
\section{Overview of the Selection}
\label{app:Selection}
Table~\ref{tab:PartDefinition} shows the definition of the objects considered in the selection as applied by \setanubis, namely, the prompt charged particles, prompt particles of any charge, and jets. The selection used matches that previously considered for the electron- and muon-coupled scenarios~\cite{ANUBIS:HNLemu2026} and is summarised in Table~\ref{tab:Selection}. 

\begin{table}[ht]
\caption{Summary of the definition of objects considered in the selections. 
The jets are reconstructed using the anti-k$_t$~\cite{Cacciari:2008gp} jet algorithm with $R=0.4$ from prompt final-state particles that pass the selections outlined below.
}
\label{tab:PartDefinition}
\begin{tabular}{ll}
\hline

\textbf{Objects} & \textbf{Selections} 
\\ 
\hline
\hline
\multirow{2}{*}{Prompt charged particles}\quad\hbox{}     & $q\neq0$, $d_\text{IP}<10~\mm$, \\
 & $|\vec p|>0.1~\GeV$, $\pt>5~\GeV$ \vspace{0.15cm}\\
\hline
\multirow{3}{*}{Prompt particles}     & $d_\text{IP}<10~\mm$, \\
& $|\vec p|>0.1~\GeV$, \\
& ID$_\text{PDG}\notin[12,14,16,18]$ \vspace{0.15cm} \\ \hline
Jets                  & $\pt>15~\GeV$                                                       \bigstrut \\ \hline
\end{tabular}
\end{table}

\newpage
\section{Summary of ANUBIS HNL limits}
\label{app:SummaryLimits}
\begin{figure}[htb]
    \centering
    \includegraphics[width=0.95\linewidth]{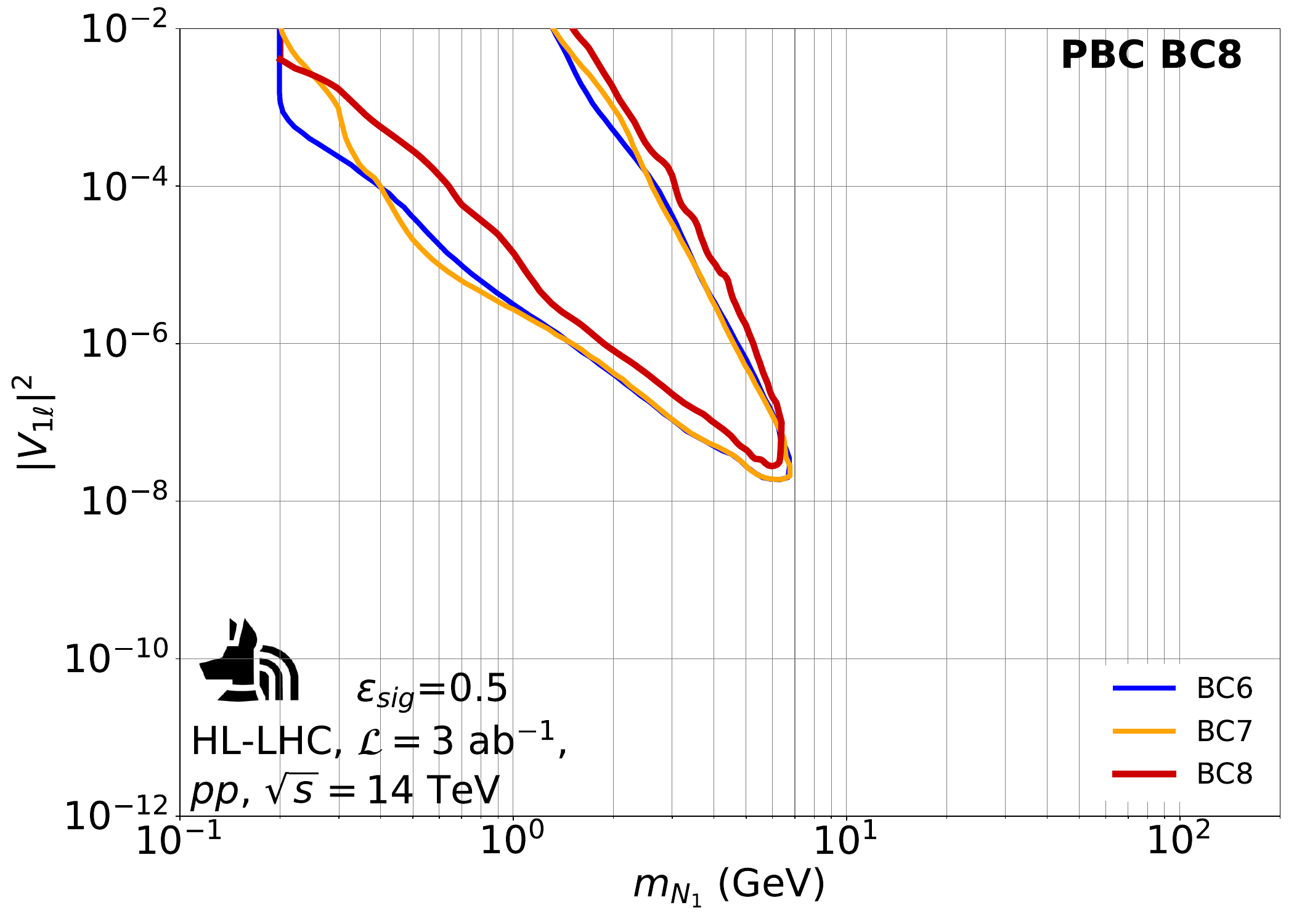}
    \caption{Projected exclusion sensitivity of the \anubis detector to a minimal Majorana HNL scenario that exclusively has either electron, muon or tau couplings, corresponding to the BC6, 7, and 8 PBC scenarios, respectively.}
    \label{fig:AllHNLLimits}
\end{figure}

Figure~\ref{fig:AllHNLLimits} shows a comparison between the sensitivity contours of BC6, 7, and 8. This highlights the impact of kinematic limitations for the BC8 case at lower HNL masses relative to the other cases. 

\onecolumngrid
\vspace{2cm}

\begin{table}[hb]
\caption{
Summary of the selection strategy implemented in \setanubis for LLP candidates to determine the geometric and kinematic acceptance. 
The $(x,y,z)$ positions of the decay vertices are determined relative to the ATLAS IP; $r_\text{CoC}$ represents the radius from the centre of curvature of the ATLAS cavern ceiling ($x=-1.7~\metre$, $y=3.52~\metre$); and $r_\text{IP}$ is the radius from the ATLAS IP. 
The selections are applied successively in the order shown.}
\label{tab:Selection}
\begin{tabular}{lll}
\hline
\textbf{Category} & \textbf{Description}      & \textbf{Requirement}                
\\ \hline 
\hline 
\multicolumn{1}{l|}{\multirow{5}{*}{\textbf{Geometric acceptance}}} &
  \multirow{2}{*}{LLP decay within the ATLAS cavern} &
  $-17.2<x<13.8~\metre$,  $y > -11.37~\metre$, $|z| < 26.4~\metre$,\\ 
\multicolumn{1}{c|}{} & & $ r_\text{CoC} < 20~\metre$ \\
\multicolumn{1}{c|}{} & ATLAS veto               &  Not within ($|z|<22~\metre$ and $r_\text{IP} < 9.5~\metre$)                  \\
\multicolumn{1}{c|}{} & LLP-ANUBIS intersection       & $N_\text{inters.} \geq 2$ layers                     \\
\multicolumn{1}{c|}{} & LLP Decay charged-track multiplicity              & $N_\text{tracks}\geq2$, $N_\text{hit} \geq 2$ layers per track \\ \hline
\multicolumn{1}{l|}{\multirow{4}{*}{\textbf{Kinematic acceptance}}} &
LLP momentum & $|\vec p_\text{LLP}| > 0.1$~\GeV\\
\multicolumn{1}{c|}{} & \met requirement               & $\met > 30~\GeV$                                \\
\multicolumn{1}{c|}{} & LLP isolation from jets              & $\Delta R_\text{LLP,jet} > 0.5 $                           \\
\multicolumn{1}{c|}{} & LLP isolation from energetic tracks & $\Delta R_\text{LLP,trk} > 0.5$                        \\ \hline
\end{tabular}
\end{table}

\end{document}